# Data-Driven Animation Controller: A Prioritized Visual System for Decoupled Animation Logic in Godot Game Engine

1st Abtin TorabNezhad
*dept. of computer Engineering, Ka.c., Islamic Azad University, Karaj, Iran*
abtin.torabnzhad@gmail.com

2nd Azam Bastanfard*
*dept. of computer Engineering, Ka.c., Islamic Azad University, Karaj, Iran*
bastanfard@iau.ac.ir

3rd Ashkan Rezaei
*dept. of computer Engineering, Ka.c., Islamic Azad University, Karaj, Iran*
ashkan.rezaei0702@iau.ir

***Abstract— This paper introduces the Data-Driven Animation Controller (DDAC), a specialized Godot component that achieves robust decoupling of animation logic from core gameplay scripts through a data-driven approach. Animation control is typically centralized and imperatively defined within core character scripts, often relying on implicit Finite State Machines (FSMs). This practice leads to tightly coupled and difficult-to-maintain codebases. The DDAC component externalizes these instructions into easily inspector-editable resources, effectively making the animation logic declarative. Rules are defined by reading Conditions from any variable on any external node and executing Actions (setting the target animation). The DDAC also manages secondary visual state settings, such as Animation Speed Scaling and Horizontal/Vertical Sprite Flipping, using the same simple rule-based setup. The highest contribution of this work is the use of a Prioritized Resolution Algorithm to enforce mutual exclusion, ensuring that when multiple rules match, only the highest-priority rule executes. This framework allows designers to quickly iterate on character-state visualization without modifying code, while significantly improving maintainability and reducing cognitive load on core developers.***

## I. Introduction

Code decoupling in game engines has recently become an important research topic because code complexity and modularity requirements of contemporary game development have increased. The evolution of game AI and behavior systems has revealed an obvious tendency for the architecture to become more modular and reusable, going beyond tight state-machine connections or behavior trees visible in initial finite-state machine construction [1] and the increasing popularity of component-based game engines [2]. Recent advances highlight the importance of behavior systems that can be integrated flexibly to enhance responsiveness and maintainability [3][4]. Faster processing has been one of the driving factors, driven by valid requirements imposed by a gaming market that generated revenues of over $60 billion as early as 2007 [5] and that requires scalable architectures [6]. Furthermore, the growing prevalence of open-source code and modular frameworks underscores the need for decoupling to support portability across platforms [7][8]. Even with modern architectures (which continue to evolve), core game logic is still usually too tightly integrated with the conditional actions used for state management. This coupling also impairs flexibility and reuse (especially in a node-based architecture like Godot). Most techniques mix the control decision with functional scripts as part of the 'if' condition itself, making it difficult to maintain and adapt [3][9].

While component-based and semantic decoupling technologies have been proposed [4][10], there are few all-encompassing, declarative solutions for conditional state mapping that designers can fully control. Architectural controversy bounds as some propose hierarchical state machines [1], others modular statecharts [11], or middleware-based decoupling [7]. However, all such methods typically require an intelligent programmer to intervene in formulating transition rules. Moreover, the lack of a standardized and explicit conflict-resolution mechanism for situations in which more than one state condition is satisfied compounds this deficiency [12][13]. This absence of a declarative conflict-resolution mechanism increases development overhead and reduces agility for game flow representation.[5].

In this paper, we propose a data-driven and modular approach for controlling many agents in competitive game with clear higher level decision and lower level execution separation. Consistent with our philosophy, it shows that decoupling behavioural logic from control mechanisms creates a more scalable and maintainable solution along with improved iteration times when working on complex multi-agent game systems. [16]

This work demonstrates how moving control logic out into structured, data-oriented forms may help to provide greater flexibility in the system at a lower cost to developers. While domain and the abstractions are not the same (visual data annotation vs. data-driven animation), the underlying declarative and rule based control concept is almost identical to our work on a framework for managing and transitioning between visual states. [17]

The research is guided by the conceptual framework that combines the main concepts of data-driven design, modularity, and decoupling. Modularity is the decomposition of a system into replaceable parts [11][14], while decoupling focuses on reducing dependencies among components to promote reusability and maintainability [4][10]. Condition-Based State Mapping. It manages actions based on the game state, except for the node-to-function mapping logic [3][15]. This paper presents the Decoupled, Rule-Based Data-Driven Animation Controller (DDAC) for the Godot Engine. The DDAC uses a data-driven, editor-defined design to enable designers to craft full sets of conditional logic and conditions within the editor itself. Specifically, this paper:

Describes the architecture of the DDAC, which employs a flat set of rules to connect game variables with specific state-setting actions across different nodes.Introduces the Prioritized Resolution Algorithm, a novel approach that uses predefined designers' preference information for mutual exclusion and conflict resolution of rules. In this, research adds value by narrowing the gap from purely theoretical decoupling strategies to a self-explanatory, practical implementation that improves the autonomy of designers and the stability of code in contemporary game development [3][2].

Innovation:

- This system is about taking the IF (condition) THEN (visual action) logic out of code and into a reusable, data-driven system.

- In other words, an animator can easily specify behaviors, while this system prioritizes multiple visual state changes based on runtime conditions or user input. Thesis: The DDAC component provides a decoupled, design-friendly, prioritized rule-based approach to visual state management -- much less error-prone than hard-coded FSMs, while promising robustness, since it encodes the choices of main animation, speed, and sprite directionality

- This tool fills the gap by offering a method that is declarative, straightforward, and focused only on the game and animation sides of mapping game variable outcomes to visual outcomes, using a method of mutual exclusion. No method or tool offers a simpler, data-based option than traditional FSMs or BTs.

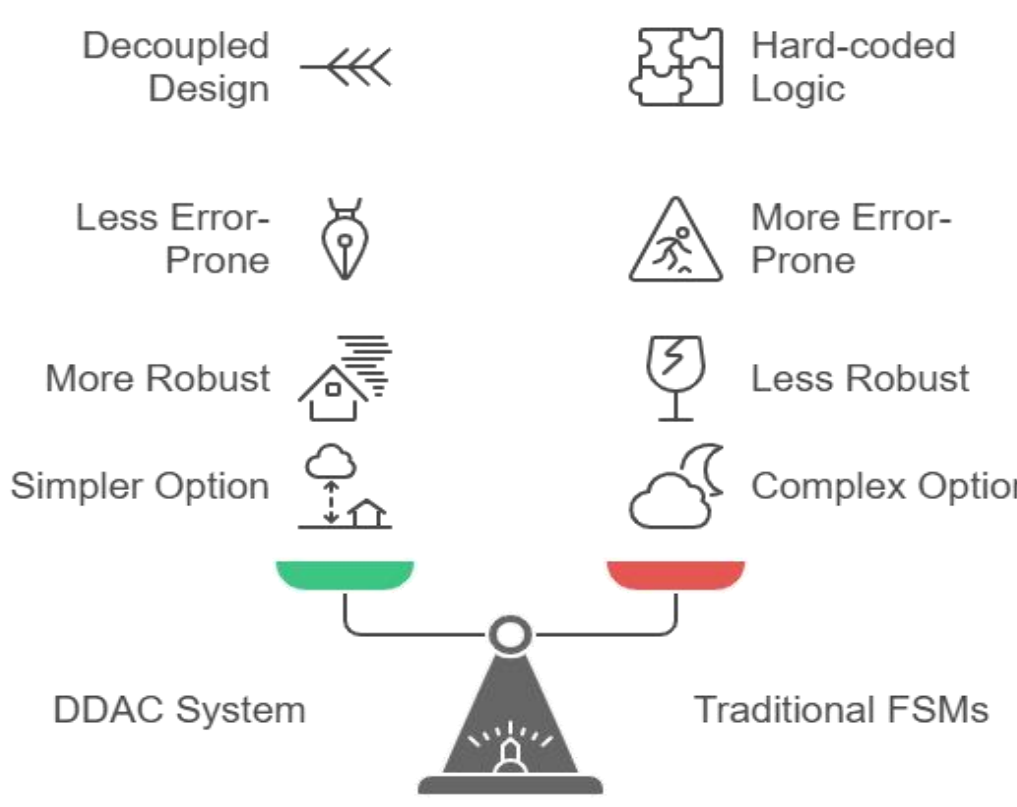

Figure 1:DDAC offers a simpler, more robust approach to visual state management.

## II. RELATED WORK

### *2.1 Animation State Machines and Engine-Level Control*

Character animation in contemporary game engines is most commonly managed through Finite State Machines (FSMs) or closely related state-based models. Tools such as Unity's Animator Controller, Unreal Engine's Animation Blueprints, and Godot's AnimationTree rely on explicit state definitions and transition rules to drive animation playback. These systems provide predictable behavior and are well suited to small or moderately complex characters. However, as character behaviors expand, FSM-based controllers tend to grow rapidly in complexity, leading to dense transition graphs and reduced clarity. In many practical implementations, animation decisions are further embedded directly within gameplay scripts, creating tight coupling between animation logic and core game behavior. This coupling complicates maintenance and makes iterative design changes increasingly costly.
In response, a range of studies has explored modular animation architectures aimed at separating animation concerns from core control logic. Approaches based on behavior trees, neural networks, and physics-driven modules enable animation components to evolve independently and integrate more easily into larger systems [18][19][20].

Other frameworks retain centralized control while allowing modular extensions, balancing scalability with architectural cohesion [21][22][23]. Despite these advances, animation decision logic in such systems often remains implicit or intertwined with execution flow, limiting true declarative separation.

*2.2 Layered and Priority-Based Animation Control*

Managing simultaneous and potentially conflicting animation requests is a persistent challenge in interactive character systems. Common solutions include layered animation structures, blending trees, and interrupt mechanisms that allow higher-priority actions—such as attacks or reactions—to override baseline locomotion. Several studies explicitly address this issue through prioritized arbitration schemes, mutual exclusion strategies, or rule-based conflict resolution mechanisms [18][24][25].

Additional work has focused on improving the efficiency and stability of these systems through techniques such as prioritized inverse kinematics, optimized search strategies, and sparse mixture models [22][26]. While effective, priority handling is often distributed across multiple subsystems or encoded implicitly within animation graphs and scripts. As a result, the resulting behavior can be difficult to reason about, particularly for designers who lack visibility into the underlying resolution process.

*2.3 Data-Driven and Visual Authoring Approaches*

To improve accessibility and support rapid iteration, game engines increasingly rely on visual and data-driven animation authoring tools. Node-based editors and visual scripting systems allow designers to define animation behavior without modifying source code, lowering the barrier to experimentation. Declarative logic interfaces and inspector-editable resources have been shown to reduce cognitive load and improve workflow efficiency [23][27][28].

Several studies extend these ideas through modular feature selection and subspace decomposition, enabling animation retrieval or synthesis to adapt dynamically to character state [25][27][29]. Visualization tools that expose gait features, behavior states, motion coverage, or training metrics further support debugging and refinement [28][30][31]. However, as visual systems grow in complexity, execution order and decision logic often become implicit, reducing transparency and making long-term maintenance more challenging.

*2.4 Decoupling Gameplay and Animation Logic*

Decoupling gameplay logic from animation control is a recurring objective in both software and game architecture. Component-based and event-driven designs aim to separate high-level behavior from visual presentation, improving modularity and reuse. Prior work has explored message broadcasting between modules, loosely coupled physics-based transitions, and event-triggered animation systems to reduce direct dependencies [32][33]. Reinforcement learning combined with imitation learning has also been investigated as a means of improving robustness and adaptability in character control [19][30].

Despite these efforts, animation logic in many systems remains fragmented across gameplay scripts, animation controllers, and engine-specific tools. This fragmentation limits designer autonomy and often necessitates programmer intervention for minor visual changes. Fully declarative, engine-integrated animation decision systems with explicit conflict resolution remain comparatively rare.

*2.5 Performance Considerations in Real-Time Animation*

Real-time performance is a fundamental requirement for interactive animation systems. Numerous studies report achieving interactive frame rates through optimizations such as level-of-detail management, efficient physics simulation, lightweight neural models, and optimized motion search techniques [20][23][32]. Algorithmic refinements, including sparse representations and prioritized computation, further contribute to maintaining responsiveness under runtime constraints [22][24].

At the same time, several works highlight trade-offs between realism and responsiveness, with systems often tailored to specific application domains such as games, simulation, or robotics [23][34]. These observations underscore the importance of animation controllers that provide predictable performance without sacrificing expressive control.

*2.6 Existing Research Gap*

Taken together, prior research highlights the importance of modularity, usability, robustness, performance efficiency, and visualization in animation control systems. However, these aspects are often addressed independently, and animation logic frequently remains tightly coupled to gameplay systems or governed by implicit and fragmented priority mechanisms. While visual and data-driven tools improve accessibility, they often lack transparent and deterministic conflict resolution. In contrast, this work introduces a data-driven, rule-based animation controller that explicitly decouples animation logic from gameplay code and employs a prioritized resolution algorithm to deterministically manage competing animation rules within the Godot game engine.

## III. Methodology

The DDAC is a dedicated Godot AnimatedSprite2D Component specifically to connect to any node to provide animated control.

The core logic is stored in a Resource Data file containing an array of rules. Each Resource data structure contains:

1. an array of Conditions (like reference value, comparison mode, path to node, variable name).

2. Priority, the principal value that is used for mutual exclusion.

3. The string name of the Animation to play.

The component has a special “Default Animation” rule with the lowest priority, which keeps a visual element in an active state, increasing stability and resilience. Horizontal Flip States, Vertical Flip States, and Animation Speed Scale. All of these use the same decoupled logic to map external variables directly to sprite variables.

The Priority-Based Algorithm (The Main Component):

At every tick, the component loops through its rules and tests their conditions. The rules whose conditions are TRUE are added to a list of Candidates. In this step, the algorithm determines the highest Priority value among the Candidate list. Only Candidates whose priority equals the highest value are kept. All lower-priority matches are Excluded Mutually. The component then executes the action from the highest priority rule. Noteworthy, if there are multiple rules with the highest priority, their actions will be run (i.e., the animations are set as well as all H-Flip states).

It should be strongly noted that this simple priority mechanism allows us to provide the cascading logic that ensures a high-priority state (i.e., Damaged: Priority 10) will always supersede a lower priority state (i.e., Running: Priority 1) without any complicated code dependencies.

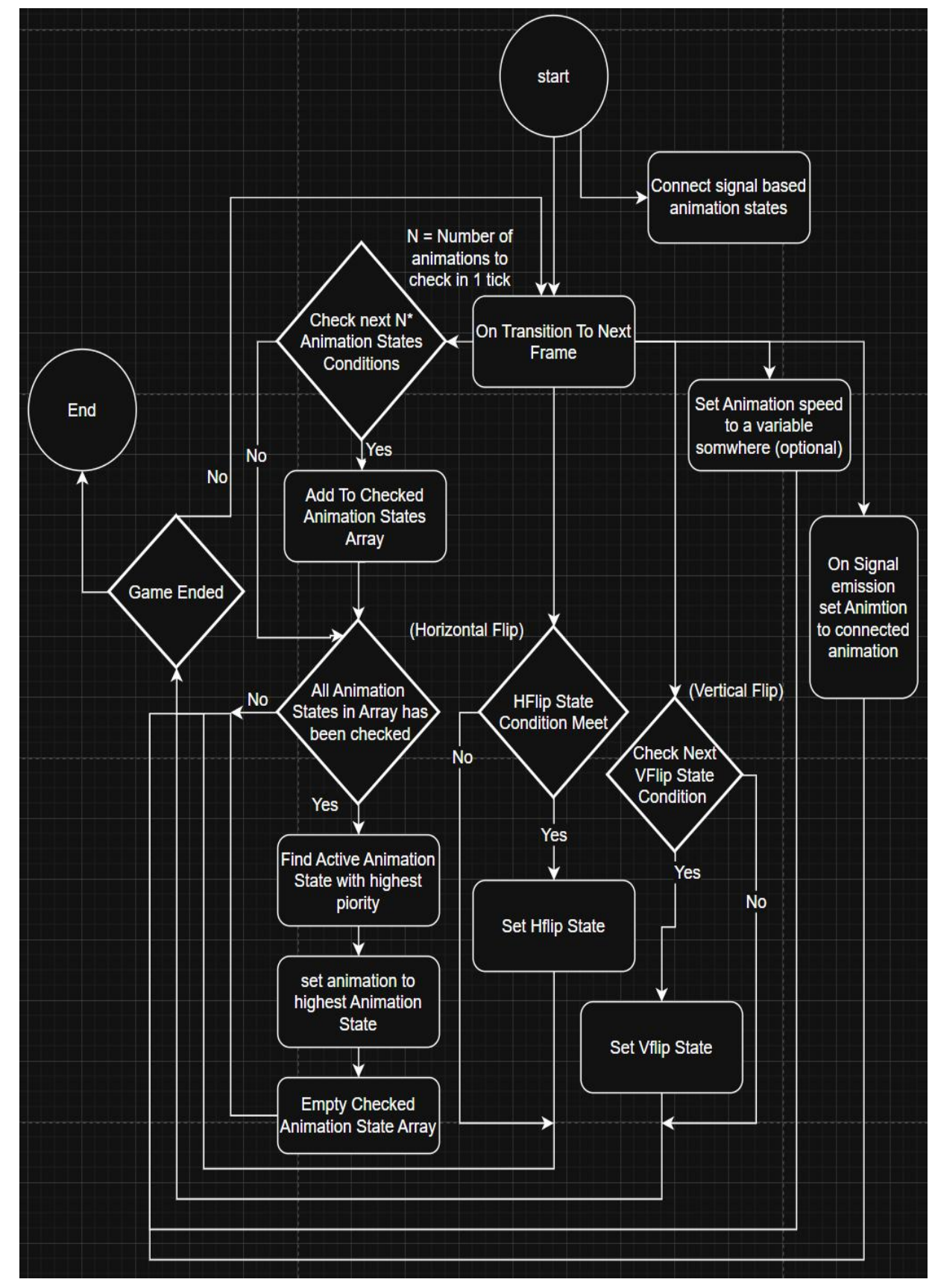

Figure2:Data-Driven Animation Controller flowchart

## IV. Results and Discussions

The Data-Driven Animation Controller, DDAC, compresses all animation code into side-effect-free, serializable resources detached from gameplay scripts. Its rule system can read state from any node and perform actions such as swapping animations, scaling speed, or flipping sprites . A Walk-the-Walk Resolution Algorithm discourages mutual exclusion to allow more than one matching rule of highest priority to execute. The proposed system can effectively make maintenance easier and allows designers to iterate quickly on character visuals without writing code.

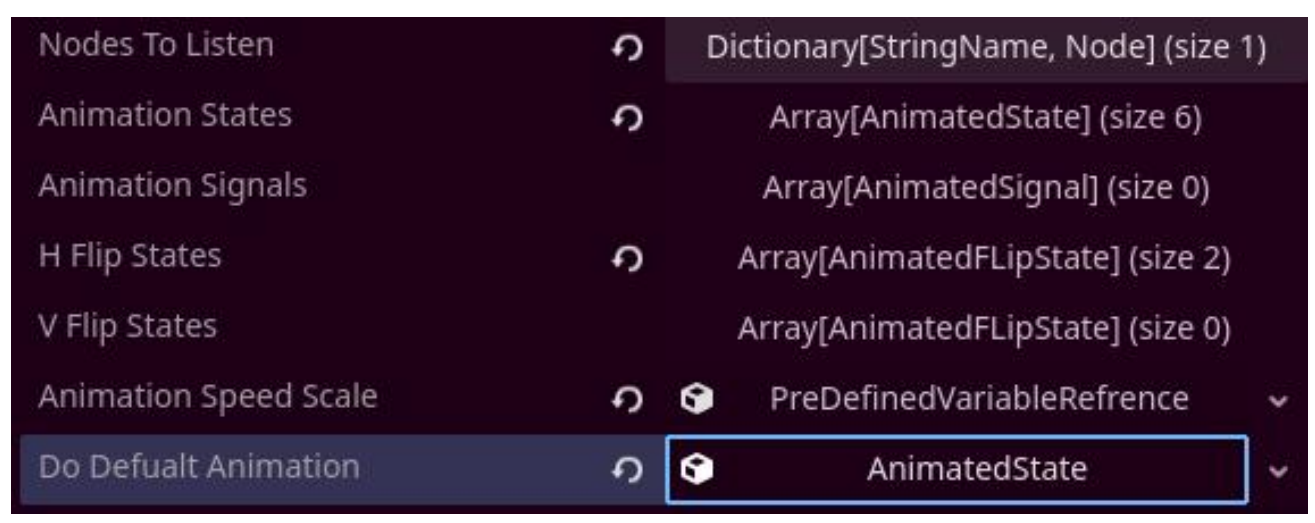

Figure 3: DDAC contains an array of nested resources, each containing changeable data and functions.

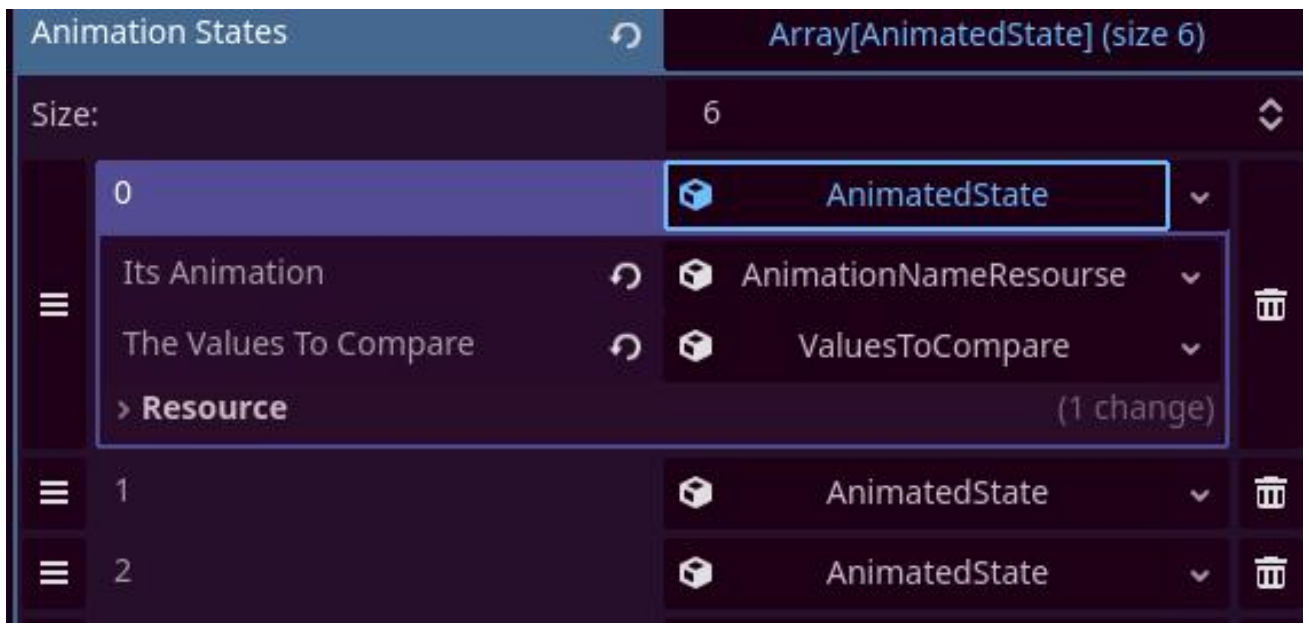

Figure 4: The most important array is the container for animation states, which all contain resources with animation and condition data.

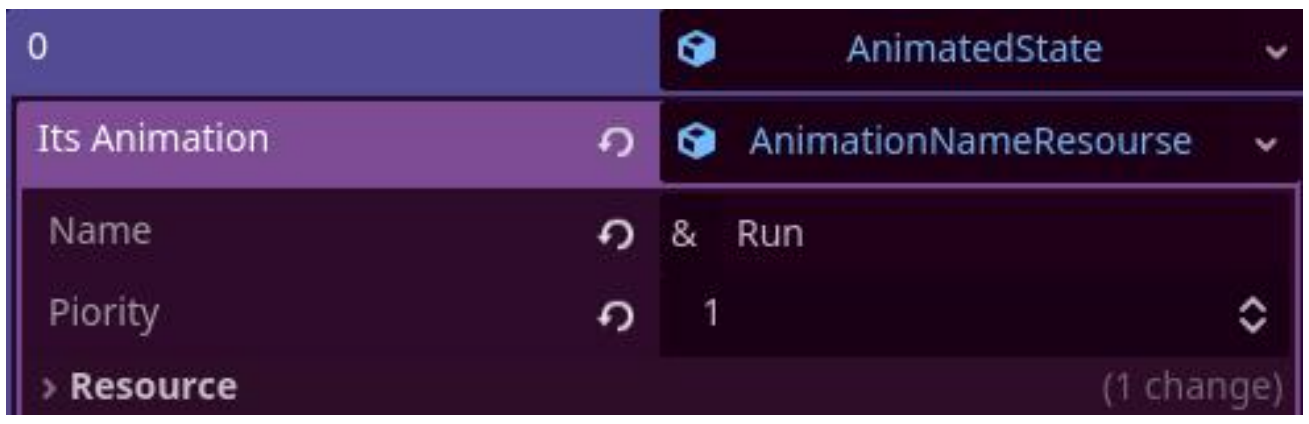

Figure 5: The AnimationNameResource contains the animation name and its priority.

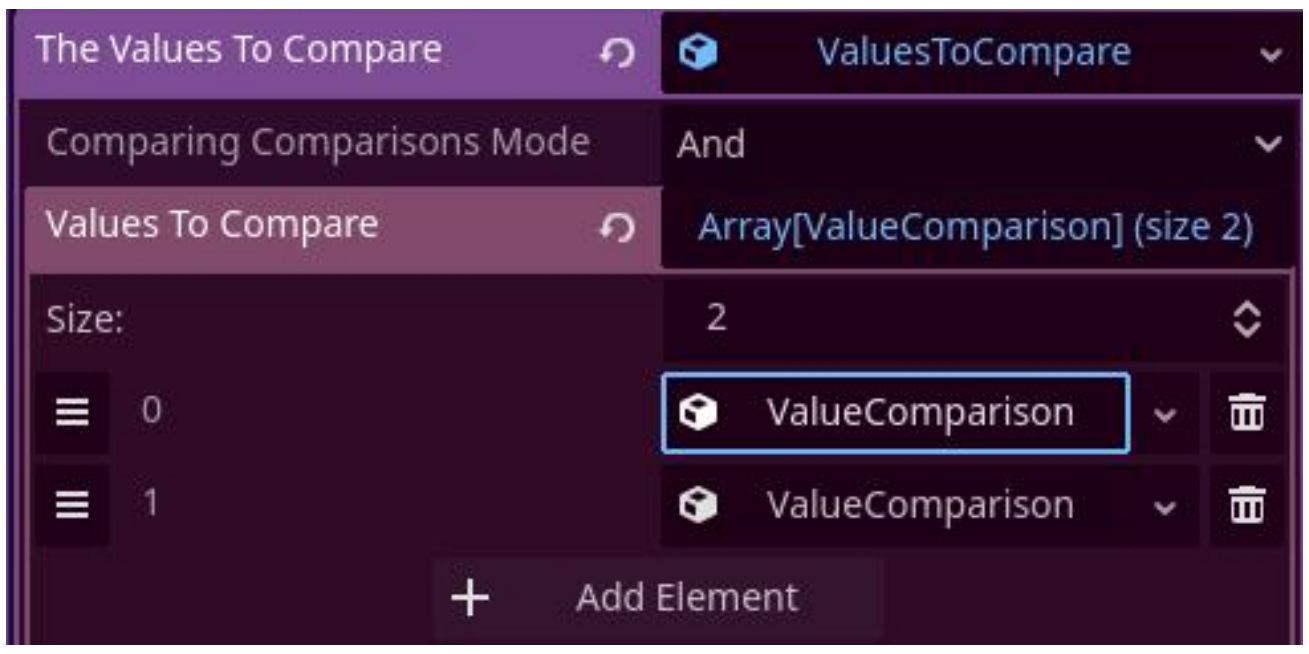

Figure 6:ValuesToCompare resource contains an array of conditions and a compression mode (And/Or)

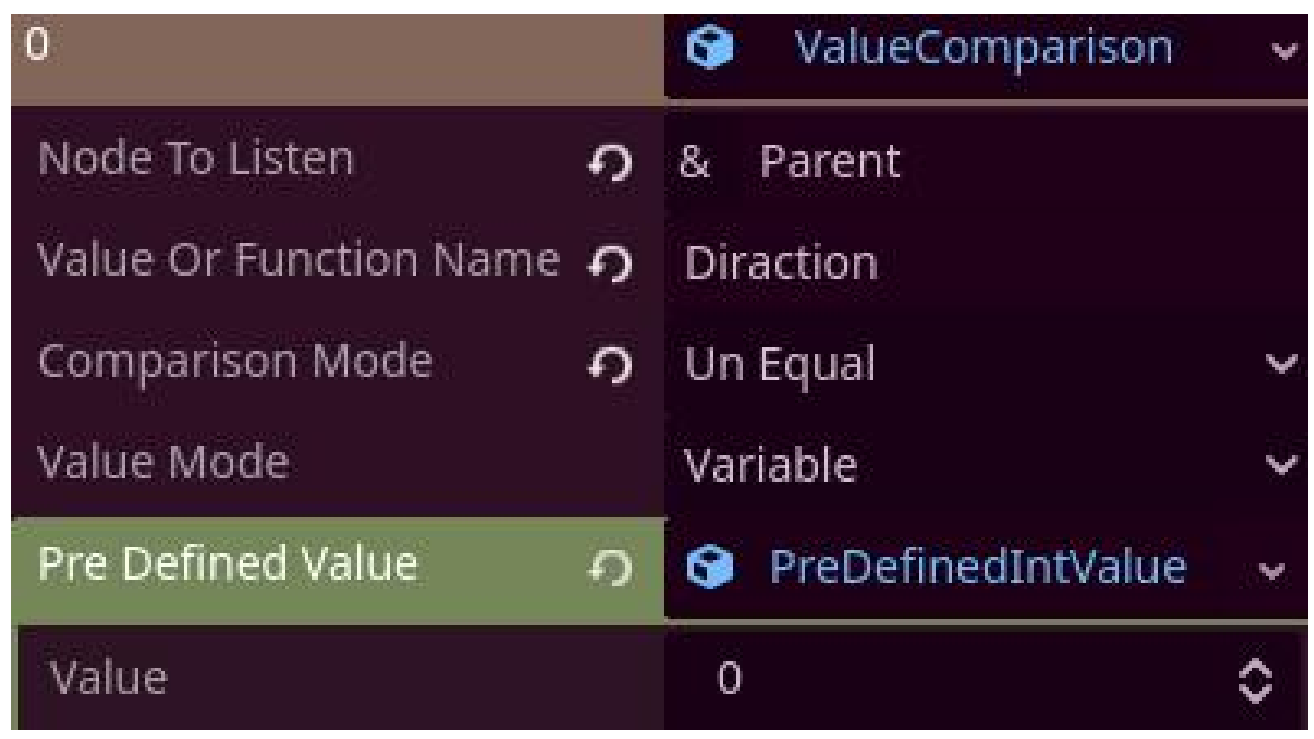

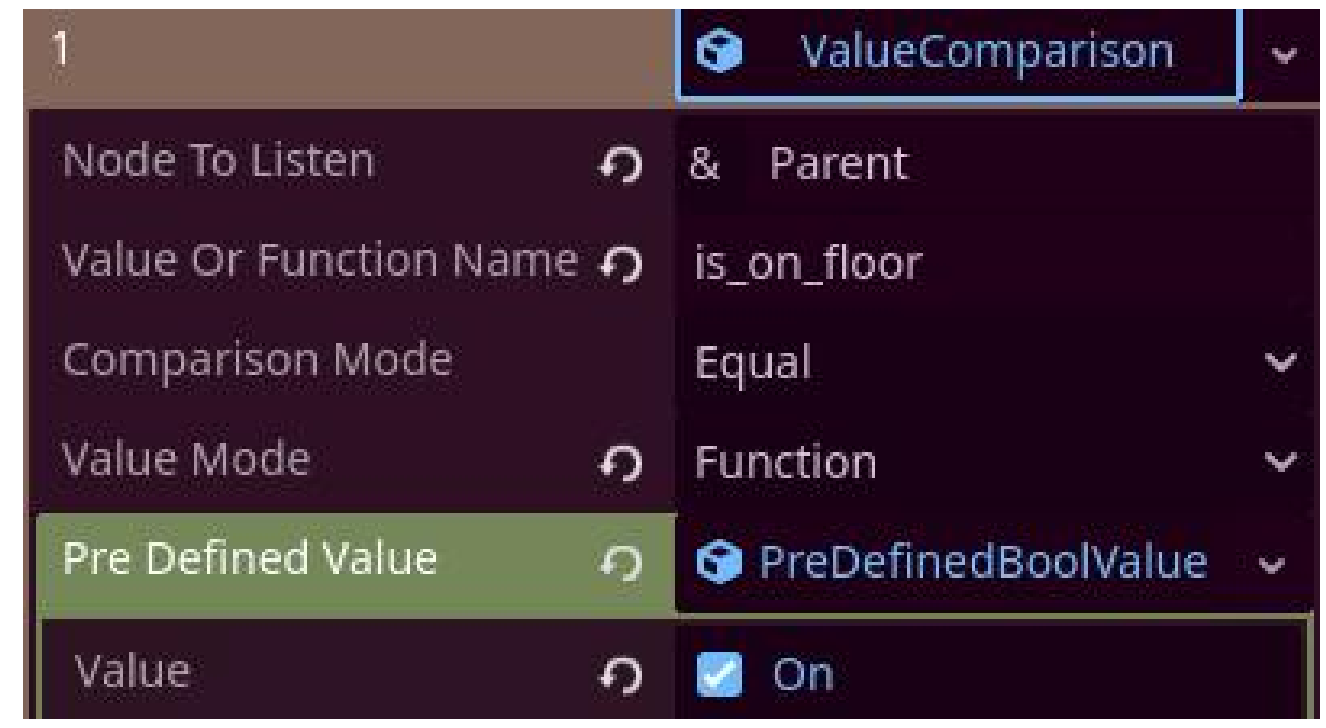

Figure 7 and Figure 8: Condition 1 includes a name referring to a user-defined path to a node, a value or function within that node, a comparison mode (Equal, Not Equal, Greater Than, Less Than), a value mode (either a static value or a function output), and a predefined value of any type.

**For example, here the conditions are:**

*Run animation rule:*

- Priority: 1
- Condition 1 : *Direction != 0* *(unequal)*
- Condition 2 : *is_on_floor == 0* *(function output)*

  *If no other rule with a higher priority is checking than the Animation will be set to the corresponding animation*

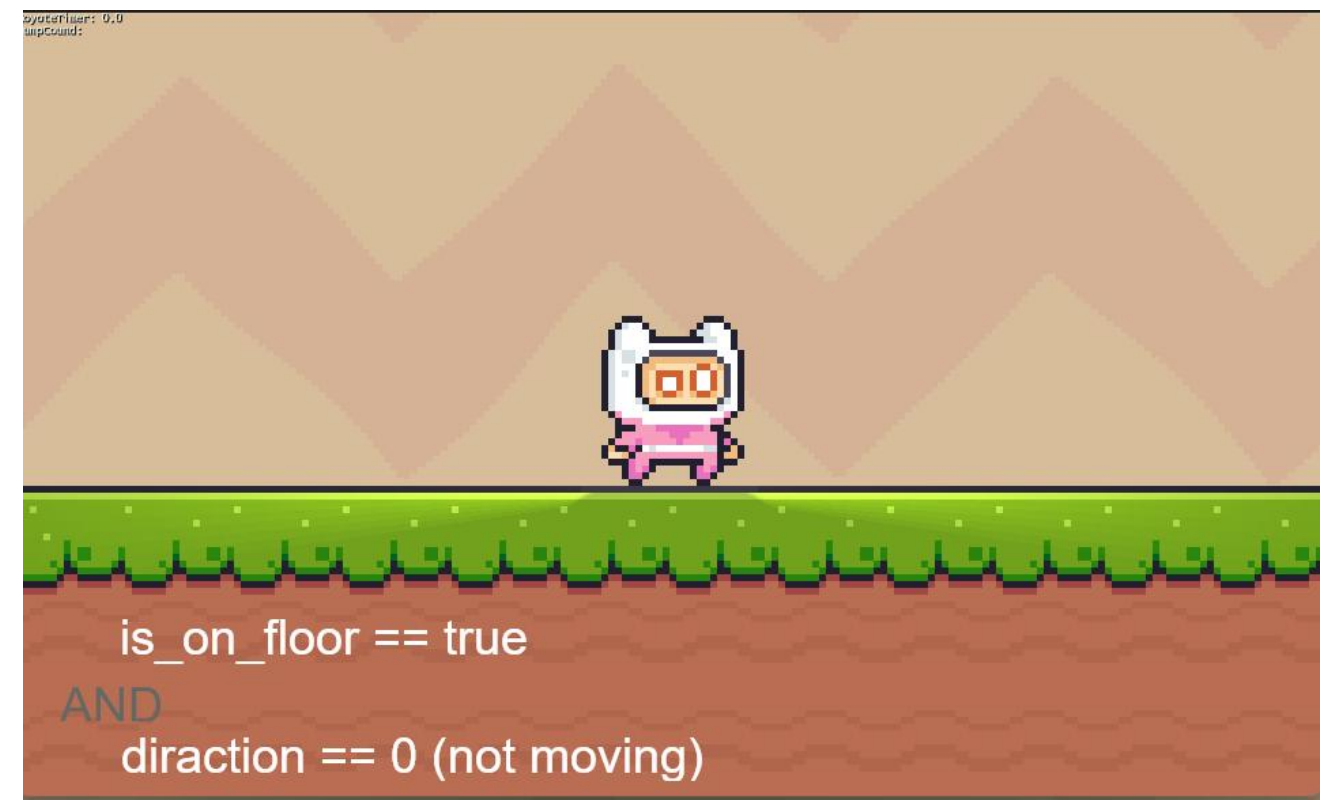

Figure 9 shows that *is_on_floor* is true and *direction* is 0.

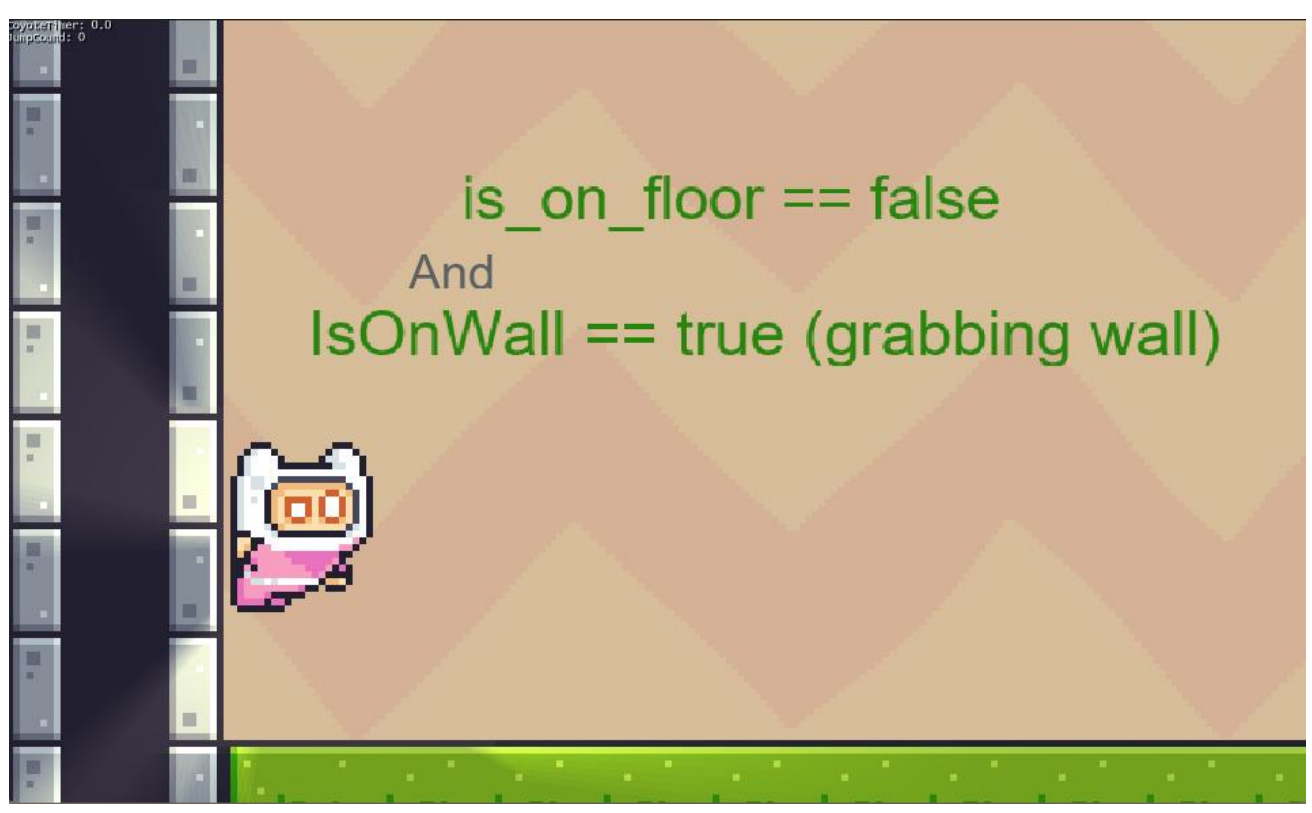

Figure 10 shows that *(is_on_floor AND IsOnWall == true)*

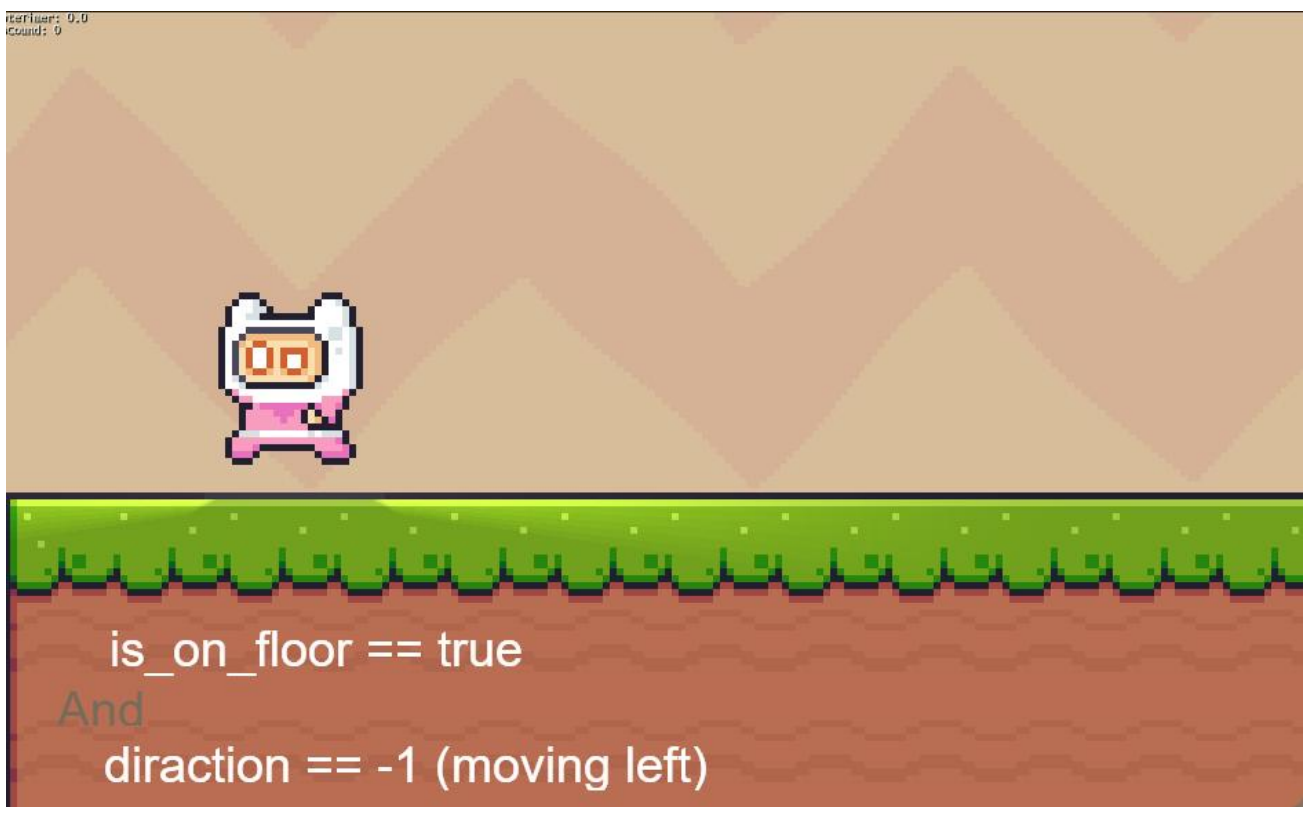

Figure 11 shows that *is_on_floor* is true and *direction* is 1 (left).

The Player (or any other node) only controls its own variables and doesn't even know the DDAC exists; the DDAC component handles the complex mapping to Animation, H_Flip, V_Flip, and AnimationSpeed by itself, decoupling the animation handling from.

Designer Empowerment: Logic is changed via data, not code.

Reduced Code Complexity: The core character script is free from animation and visual logic.

Runtime Flexibility: Rules can be added or disabled instantly without recompiling.
While visual results promise a positive effect on performance efficiency, it may be prudent to follow up with further evaluations. Evaluation methods such as saliency-based image quality evaluation [35], can be incorporated to enable more human-aligned quality assessment.

## V. Conclusion

The DDAC effectively uses a decoupled, prioritized rule-based strategy to control animations, improving maintainability and designer iteration speed by integrating all significant visual state properties (animation name, speed, and flipping).

Future Work:
Adding event-driven checking, re-evaluating rules only when one of the source variables changes, for performance rather than a constant tick loop. Designing a simple visual graph view of the rules array in the inspector. Creating a way to compare two variables or a function output to a variable, rather than only comparing a variable/function output to a predetermined value.Generalizing the whole system.